# Prediction of Binding Affinity for ErbB Inhibitors Using Deep Neural Network Model with Morgan Fingerprints as Features

La Ode Aman[1,*]

[1] Department of Pharmacy, Faculty of Sport and Health, Universitas Negeri Gorontalo, Gorontalo, Indonesia

* Correspondence author: laode_aman@ung.ac.id

Abstract

The ErbB receptor family, including EGFR and HER2, plays a crucial role in cell growth and survival and is associated with the progression of various cancers such as breast and lung cancer. In this study, we developed a deep learning model to predict the binding affinity of ErbB inhibitors using molecular fingerprints derived from SMILES representations. The SMILES representations for each ErbB inhibitor were obtained from the ChEMBL database. We first generated Morgan fingerprints from the SMILES strings and applied AutoDock Vina docking to calculate the binding affinity values. After filtering the dataset based on binding affinity, we trained a deep neural network (DNN) model to predict binding affinity values from the molecular fingerprints. The model achieved significant performance, with a Mean Squared Error (MSE) of 0.2591, Mean Absolute Error (MAE) of 0.3658, and an R-squared ($R^2$) value of 0.9389 on the training set. Although performance decreased slightly on the test set ($R^2$ = 0.7731), the model still demonstrated robust generalization capabilities. These results indicate that the deep learning approach is highly effective for predicting the binding affinity of ErbB inhibitors, offering a valuable tool for virtual screening and drug discovery.

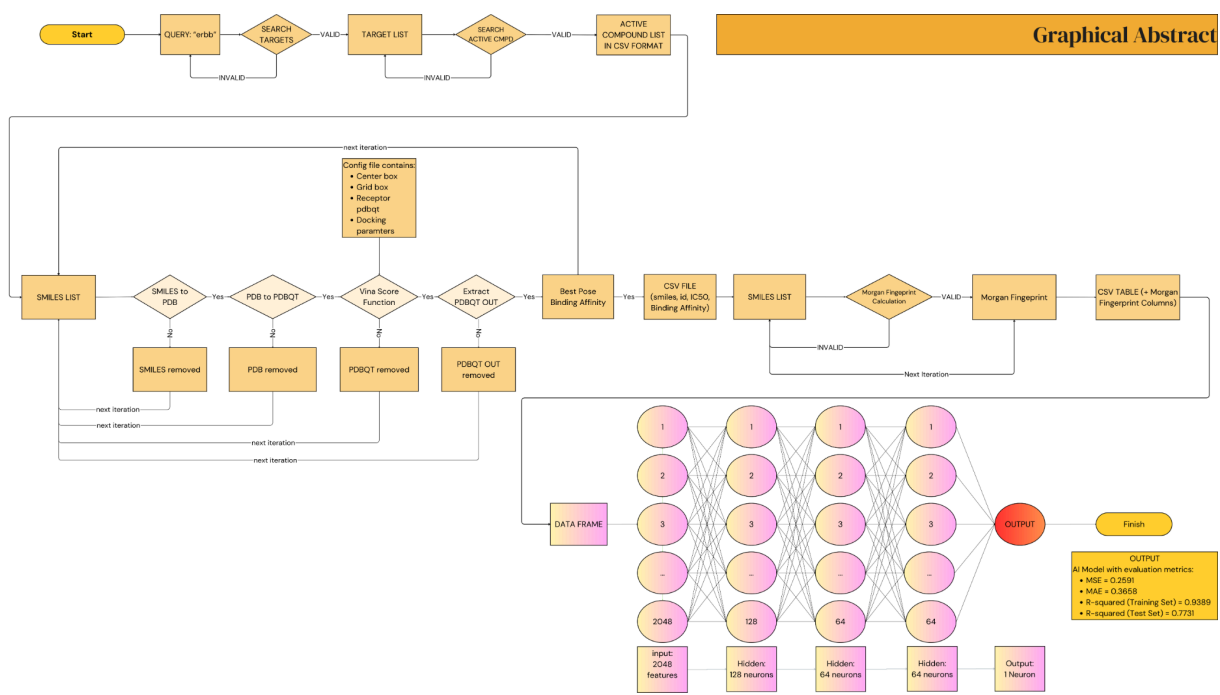

Graphical Abstract

Introduction

The ErbB receptor family, which includes ErbB1 (EGFR), ErbB2 (HER2), ErbB3, and ErbB4, is integral to various cellular processes such as growth, survival, and differentiation. These receptor tyrosine kinases (RTKs) are crucial in mediating signaling pathways that promote cell proliferation and survival, notably the PI3K-AKT and MAPK pathways [1], [2]. ErbB signaling, including overexpression and mutations, significantly contribute to the development and progression of several cancers, including breast, lung, and gastric cancers [3], [4]. For instance, overexpression of EGFR and HER2 has been linked to aggressive tumor phenotypes and poor prognosis in breast cancer [4], [5]. Specific mutations, such as the EGFR T790M mutation, can confer resistance to existing therapies, highlighting the complexity of targeting these receptors effectively [4].

ErbB receptors have emerged as pivotal drug targets in oncology due to their role in cancer progression. The development of selective and potent inhibitors for these receptors, however, remains a significant challenge. This is largely due to the molecular complexity of the ErbB signaling network, which includes compensatory activation of other ErbB family members that can sustain pro-survival signaling pathways [6]. For example, the pan-ErbB inhibitor dacomitinib has shown promise in overcoming resistance mechanisms in various cancer models, including head and neck and ovarian cancers [3], [6]. Despite these advancements, off-target effects and the intricate interplay between different ErbB receptors complicate the therapeutic landscape [7].

The therapeutic implications of targeting ErbB receptors are underscored by the development of inhibitors that can block multiple receptors simultaneously. For instance, studies have demonstrated that dual inhibition of EGFR and HER2 can significantly enhance treatment efficacy in breast cancer cells harboring specific mutations [4]. Additionally, the role of ErbB3 and ErbB4 in signaling redundancy and resistance mechanisms further complicates treatment strategies, as these receptors can contribute to therapeutic failure when targeted in isolation [5]. As such, ongoing research is focused on elucidating the precise roles of these receptors in cancer biology to inform the design of more effective combination therapies [6], [8].

Current therapeutic strategies for targeting ErbB receptors encompass a variety of approaches, including tyrosine kinase inhibitors (TKIs), monoclonal antibodies, and antibody-drug conjugates (ADCs). TKIs, such as gefitinib and erlotinib, function by binding to the ATP-binding site of the epidermal growth factor receptor (EGFR), effectively inhibiting its kinase activity and downstream signaling pathways [9]. Advanced versions of these inhibitors, like osimertinib, have been specifically designed to overcome resistance mutations, particularly the T790M mutation in EGFR, which is a common mechanism of resistance in non-small cell lung cancer (NSCLC) [10].

Monoclonal antibodies, such as trastuzumab and pertuzumab, target HER2 to block its activation and prevent receptor dimerization, which is critical for the oncogenic signaling mediated by this receptor [11], [12]. Trastuzumab binds to the extracellular domain of HER2, while pertuzumab targets a different domain, providing a complementary mechanism of action that enhances therapeutic efficacy in HER2-positive breast cancer [12]. This dual targeting has shown improved outcomes in patients with HER2-positive tumors, demonstrating the importance of these antibodies in the treatment landscape [11], [12].

ADCs represent an innovative therapeutic strategy that combines the specificity of monoclonal antibodies with the cytotoxicity of chemotherapeutic agents. These conjugates are designed to selectively deliver cytotoxic drugs directly to tumor cells expressing ErbB receptors, thereby minimizing off-target effects and enhancing therapeutic efficacy [13]. The use of ADCs targeting ErbB receptors has shown promise in preclinical and clinical settings, indicating their potential as effective treatments for various cancers [13].

Despite advancements in therapeutic strategies targeting ErbB receptors, significant challenges persist in developing effective inhibitors. One major issue is the high homology within the ATP-binding domains of the ErbB family members, which complicates the design of selective inhibitors. This homology often leads to cross-reactivity and reduced specificity among inhibitors, making it difficult to achieve the desired therapeutic effects without affecting other kinases Niggenaber et al. [14]. Additionally, tumors frequently develop resistance through mutations, such as the EGFR T790M mutation, or by activating compensatory signaling pathways, which further diminishes the therapeutic efficacy of these inhibitors [15].

Off-target effects, including skin rashes and gastrointestinal disturbances, also limit the therapeutic window of ErbB inhibitors, impacting patient quality of life [16]. The unique characteristics of ErbB3 and ErbB4 present additional challenges; for instance, ErbB3 lacks intrinsic kinase activity, necessitating alternative strategies to disrupt its interactions with other receptors [17]. This lack of intrinsic activity complicates the direct inhibition of ErbB3, requiring innovative approaches to target its role in signaling networks effectively [18].

To address these challenges, advances in structural biology have facilitated the adoption of structure-based drug design (SBDD). This approach allows for precise modeling of protein-ligand interactions, enabling the development of drugs with enhanced specificity and potency [19]. Furthermore, the use of allosteric inhibitors, which target regions outside the ATP-binding pocket, offers a promising strategy to circumvent resistance issues by focusing on alternative binding sites [20].

Combination therapies have also emerged as a viable solution to resistance mechanisms. Pairing ErbB inhibitors with agents that target parallel pathways, such as mTOR or MEK inhibitors, can effectively mitigate compensatory signaling and improve treatment outcomes [21]. Additionally, the integration of artificial intelligence (AI) and machine learning into drug discovery processes has revolutionized the field, facilitating the prediction of ligand binding and optimization of candidate molecules, thereby accelerating the development of effective inhibitors [22].

Finally, the rise of patient-specific therapies represents a significant advancement in oncology. By leveraging genomic profiling and biomarker identification, personalized treatment strategies can be designed to address the unique molecular characteristics of individual tumors. This tailored approach not only enhances therapeutic efficacy but also minimizes off-target effects, providing a more precise and patient-centric form of cancer treatment [23].

In recent years, computational methods, particularly machine learning (ML) and deep learning (DL), have emerged as promising approaches to accelerate the drug discovery process. These methods offer the advantage of predicting molecular properties, including binding affinity, with relatively low computational cost compared to traditional experimental techniques. By leveraging molecular representations like SMILES (Simplified Molecular Input Line Entry System) strings and fingerprint-based descriptors, machine learning models can be trained to predict how well a compound interacts with specific targets, such as the ErbB receptor [24], [25].

Machine learning approaches have been successfully implemented for predicting binding affinity, utilizing various features of molecular complexes. For instance, deep learning frameworks have been employed to analyze atomic-level features and protein sequence-level characteristics to enhance the accuracy of binding affinity predictions [25]. Additionally, the integration of multi-omics data with deep learning has shown potential in predicting drug responses in cancer, further demonstrating the versatility of these computational methods in drug discovery [26].

The application of machine learning methods extends to the development of empirical scoring functions that predict protein-ligand binding affinity. This approach has been supported by studies demonstrating the superior predictive performance of targeted scoring functions, which integrate computational systems biology with machine learning techniques to address protein-ligand interactions effectively [27]. Moreover, deep learning models have been utilized to predict drug-target interactions (DTIs) by analyzing chemical sequences and protein sequences, showcasing their capability to identify potential drug candidates without the need for structural information [28].

Recent advancements in deep learning have also facilitated the exploration of chemical space and the prediction of molecular properties, which are crucial for drug discovery. For example, frameworks that utilize graph neural networks have been developed to predict binding affinities based on intermolecular contacts, highlighting the effectiveness of these models in capturing complex molecular interactions [29]. Furthermore, the integration of AI and machine learning into drug discovery processes has revolutionized the field, allowing for the rapid identification of potential drug candidates and the optimization of lead compounds [30], [31].

In this work, we aim to predict the binding affinity of ErbB inhibitors using Morgan fingerprints, a type of molecular fingerprint that captures important structural information from SMILES representations. The model developed in this study utilizes a deep neural network (DNN) to predict binding affinities from molecular features, allowing for more efficient drug discovery and virtual screening of potential ErbB inhibitors. The study focuses on processing a dataset of 9,947 ligands, employing AutoDock Vina for initial docking calculations, and subsequently using the generated molecular fingerprints as features for model training [32].

Our approach aims to demonstrate the potential of using deep learning in combination with molecular fingerprints to predict binding affinities, providing valuable insights into the design of selective and potent ErbB inhibitors for therapeutic use. The results from this study contribute to the growing body of research that seeks to integrate computational approaches with experimental drug discovery, ultimately enhancing the efficiency and success rate of developing new therapeutic agents [33], [34].

Morgan fingerprints, known for their ability to represent molecular structures effectively, have been widely used in cheminformatics. They allow for the encoding of molecular features that can be utilized in machine learning models to predict various properties, including binding affinity [32]. The integration of deep learning techniques with these fingerprints has shown promise in improving prediction accuracy, as evidenced by recent studies that leverage deep neural networks for binding affinity predictions [34], [35].

Furthermore, the use of AutoDock Vina for initial docking calculations provides a robust framework for assessing ligand-receptor interactions, which can then be refined through machine learning models trained on the generated molecular fingerprints [36]. This combination of computational docking and machine learning represents a significant advancement in the field of drug discovery, particularly for targeting specific receptors like ErbB [37].

The search for macromolecules was conducted in the ChEMBL database, which is a comprehensive resource for bioactivity data related to drug discovery. ChEMBL provides a wealth of information on small molecules and their biological activities, making it an essential tool for researchers in medicinal chemistry and pharmacology. One relevant reference is the work by Gaulton et al., which discusses the ChEMBL database as a large-scale bioactivity database for drug discovery. This article highlights the database's broad coverage across various targets, organisms, and bioactivity measurements, as well as its user-friendly search capabilities Gaulton et al. [38]. Additionally, Liu et al. provide insights into the data curation approaches adopted by ChEMBL, emphasizing the selective range of scientific journals considered for collecting bioassay data, including protein-ligand binding data [39]. Moreover, the study by Papadatos et al. focuses on the activity, assay, and target data curation within the ChEMBL database, detailing the extensive bioactivity data available for researchers [40]. Another important reference is the work by Bento et al., which discusses the ChEMBL bioactivity database and its integration with other key public databases, providing a comprehensive overview of the data extraction and curation processes [41].

Furthermore, the article by Willighagen et al. explores the ChEMBL database as linked open data, demonstrating its integration with cheminformatics and the potential for enhanced data accessibility [42]. Lastly, the research by Kalliokoski et al. presents a statistical analysis of mixed IC50 data, illustrating the importance of accurate data representation within the ChEMBL database [43].

The 3D structure of the ErbB4 protein was generated using homology modeling with the Swiss-Model tool, accessible at https://swissmodel.expasy.org. The Swiss-Model tool is widely recognized for its efficiency in generating high-quality protein models through homology modeling. Waterhouse et al. detail the capabilities of Swiss-Model, emphasizing its optimized performance for comparative protein modeling and the generation of reliable models for various research applications [44]. Additionally, the work by Bienert et al. discusses the features and functionalities of the SWISS-MODEL repository, which provides access to pre-computed homology models and facilitates the exploration of protein structure space [45]. The use of the ErbB2 receptor (UniProt ID P04626) as a template is supported by studies that highlight the structural similarities among the ErbB family members, which can be leveraged to model related proteins such as ErbB4 [46]. The importance of evaluating model quality is underscored by Sharma et al., who discuss the evaluation metrics used in homology modeling, including Global Model Quality Estimate (GMQE) scores, which help in selecting the best model from multiple generated candidates [47]. The integration of ligands in the modeling process, as seen with the inclusion of the FMM ligand, is essential for accurately predicting the binding interactions and functional implications of the modeled protein. This aspect is highlighted in studies that emphasize the significance of ligand presence in achieving realistic structural representations [48].

Methods

*Hardware and Software Configuration*

To ensure efficient execution of computational tasks in this study, a high-performance workstation and state-of-the-art software tools were employed for data preprocessing, molecular docking simulations, and deep learning model development.

*Hardware Configuration*

All computational processes, except for molecular docking simulations, were carried out on a Hewlett-Packard HP Z840 workstation with advanced specifications. This workstation is equipped with an Intel® Xeon® E5-2650 v3 processor, featuring 40 cores, which enables efficient parallel processing with robust multi-threading capabilities, essential for handling intensive computational tasks. It also has 32.0 GiB of RAM, providing ample capacity to manage large datasets and memory-demanding applications. For deep learning computations, the workstation is outfitted with an NVIDIA GeForce RTX™ 3060 graphics card, specifically optimized to support deep learning-related calculations. Additionally, the system boasts a 3.0 TB storage capacity, sufficient to accommodate large datasets and intermediate computational results.

However, for molecular docking simulations, the Fugaku Supercomputer was utilized, leveraging its exceptional computational power to perform precise and efficient calculations of ligand-receptor binding affinities. Fugaku's unparalleled processing capabilities ensure accurate results, making it ideal for complex molecular simulations.

*Software Environment*

The software environment for the project was carefully configured with a range of tools to support various computational tasks. The operating system used was Ubuntu 24.10 (64-bit), chosen for its stability and reliability as a platform for scientific computing. It ran on the Linux 6.11.0-9-generic kernel, which ensured optimal performance for complex computations. For the user interface, the GNOME 47 desktop environment with the Wayland windowing system was employed, providing a modern and responsive interface for efficient navigation and control.

Several key libraries were integral to the project's success. RDKit (Version 2024.09.1) was utilized for cheminformatics tasks, particularly for generating molecular fingerprints and handling SMILES representations. Open Babel (Version 3.1.1) played a crucial role in facilitating format conversions, such as transforming SMILES into PDBQT format. TensorFlow (Version 2.13) was used to construct and train deep learning models designed to predict ligand-receptor binding affinities. For data preprocessing, normalization, and model evaluation, Scikit-learn (Version 1.4.1) was employed, ensuring the quality and consistency of the data. Pandas (Version 2.2.0) was indispensable for data manipulation, transformation, and in-depth analysis throughout the project.

In terms of molecular docking simulations, AutoDock Vina (Version 1.2.5) was the software of choice, run on the Fugaku Supercomputer. It was used to simulate ligand-receptor interactions and to calculate binding affinities with high precision.

For data visualization and result analysis, the project relied on several powerful libraries. Matplotlib (Version 3.8.1) was used to generate scatter plots and other types of performance visualizations, while Seaborn (Version 0.13.0) helped create more detailed and visually appealing graphical representations of data trends, making the results easier to interpret and present.

**Searching Macromolecules with Keywords**
The search for macromolecules was conducted in the ChEMBL database using the keyword "erbb", which is associated with the epidermal growth factor receptor (EGFR) family. This query identified protein targets related to the ErbB receptor family, including EGFR (ErbB1), HER2 (ErbB2), HER3 (ErbB3), and HER4 (ErbB4).
The results from this search, including the unique identifiers (target_chembl_id) and preferred names (pref_name), were used in subsequent steps to search for small molecule inhibitors. The Python code used for this step is provided in Appendix 1.

**Searching Small Molecules as ErbB Inhibitors**

Using the target_chembl_id entries obtained from the macromolecule search, small molecule inhibitors were retrieved from the ChEMBL database. The dataset was refined to include only compounds with available IC50 (half-maximal inhibitory concentration) values, ensuring that bioactivity data were present for all selected molecules.

IC50 was employed as a potency metric, with lower values indicating stronger inhibitory activity. Molecules with missing IC50 data or IC50 values exceeding a predefined threshold were excluded to focus on potent inhibitors.

This step resulted in a curated dataset of small molecules targeting ErbB receptors, prepared for further computational workflows, including data cleaning, molecular docking, and model development. The Python implementation for this process is available in Appendix 2.

**Cleaning Data**

To enhance the quality and reliability of the dataset, a systematic data-cleaning process was carried out. First, duplicate entries were addressed by consolidating molecules with identical molecule_chembl_id. Among the duplicates, only the entry with the lowest standard_value, indicative of higher potency, was retained to ensure that the most relevant data were preserved.

Next, rows containing missing or invalid values for standard_value, such as those marked as NaN or assigned a value of zero, were removed. This step was essential to eliminate incomplete or erroneous data that could compromise the analysis.

Finally, columns deemed irrelevant to subsequent computational workflows, such as pref_name and search_term, were excluded from the dataset. These columns did not contribute meaningful information to the predictive modeling process and were removed to streamline the dataset for further analysis.

The implementation of these data-cleaning procedures, designed to ensure data integrity and relevance, is documented in Appendix 3.

**Binding Affinity Calculation Using AutoDock Vina**

Binding affinities were predicted using AutoDock Vina, a molecular docking tool that simulates the interactions between small molecules and macromolecular targets. The process involved the preparation of both the receptor (macromolecule) and ligands (small molecules).

*Receptor Preparation*

The protein structures were optimized to ensure structural integrity and functionality for docking simulations. Binding pockets, representing active sites where native ligands interact, were defined using pre-specified grid box coordinates to confine the docking simulations to biologically relevant regions.

The 3D structure of the ErbB4 protein was generated using homology modeling with the Swiss-Model tool, accessible at https://swissmodel.expasy.org. The homology modeling template used was the protein with UniProt ID P04626, corresponding to the ErbB2 receptor tyrosine kinase, with FMM as the reference ligand. Among the generated models, the fifth model was selected based on the highest GMQE score of 0.74. This model was chosen as it included the FMM native ligand and corresponded to the receptor tyrosine-protein kinase ErbB4.

The binding site validation was carried out using a re-docking protocol involving the native ligand, as described in reference [49]. The docking process utilized the Vina scoring function, configured with specific parameters to ensure accuracy and reliability. The grid center was positioned at coordinates X = -37.005, Y = 50.614, and Z = 14.953, while the grid size was set to X = 40, Y = 40, and Z = 40, providing sufficient coverage of the binding site. A grid spacing of 0.375 ensured precise sampling of the search space. To enhance the thoroughness of the docking calculations, the exhaustiveness parameter was set to 8, and the computational task was executed on 32 CPUs for optimal performance.

The best docking pose yielded a binding affinity of -10.09 kcal/mol. The overlap between the docked pose (purple carbon) and the initial native ligand conformation (green carbon) in the receptor's binding pocket is displayed in **Figure 1**.

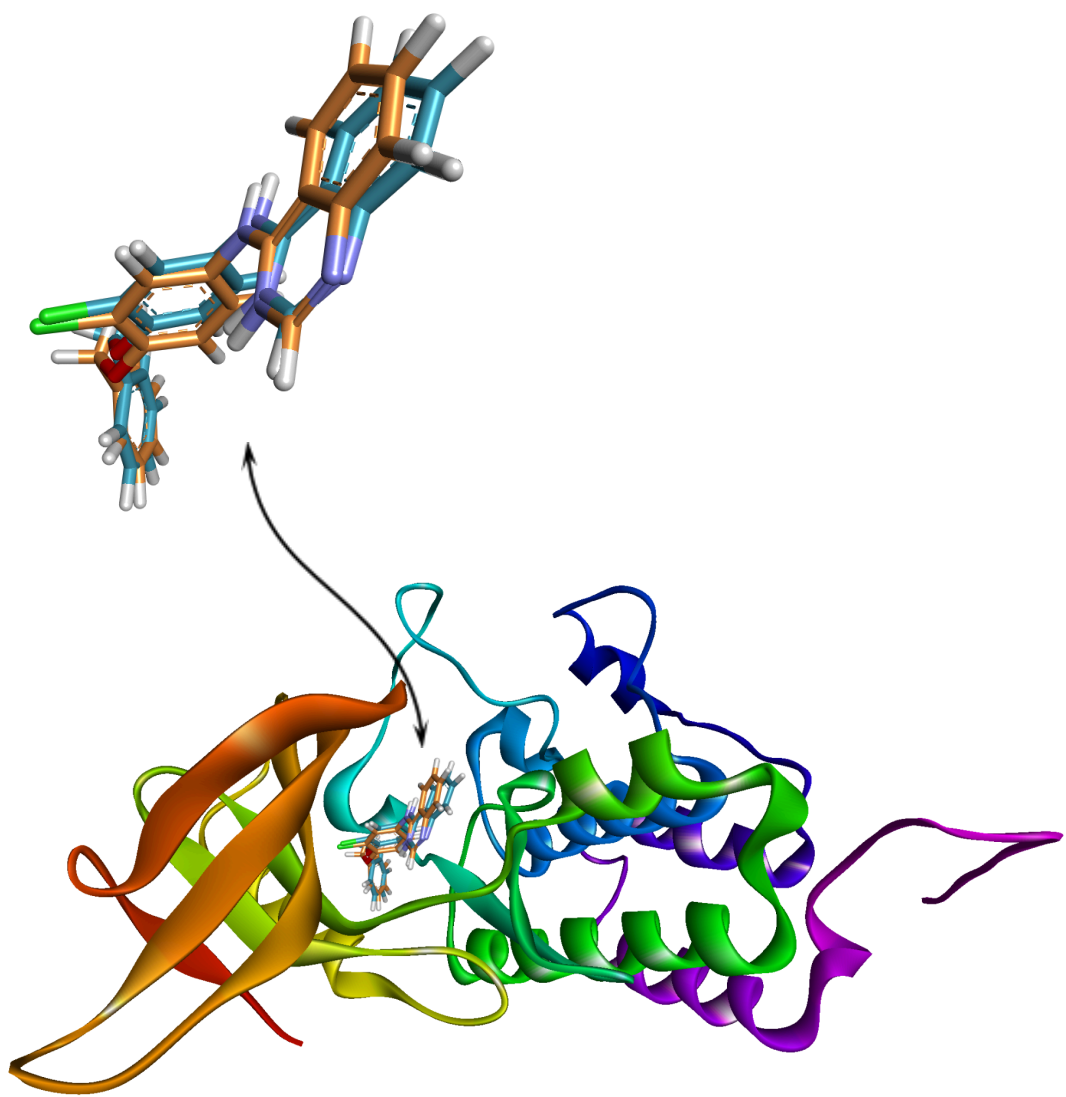

**Figure 1.** Overlay of the docked native ligand (orange carbon) with its initial conformation (blue carbon) within the receptor's binding pocket.

*Ligand Test Docking Protocol*

Each small molecule was systematically docked into the receptor's binding site using the same protocol as that employed for the re-docking of the native ligand. The docking simulation calculated the binding affinity for each interaction, expressed in kcal/mol. Molecules with binding affinities below -5 kcal/mol were considered to have significant potential as inhibitors and were selected for further analysis. Details of the configuration parameters and docking scripts used in this process are provided in Appendix 4.

**Morgan Fingerprint Calculation**

To numerically represent the structural characteristics of small molecules, Morgan fingerprints were generated using the RDKit library. These fingerprints are binary vectors encoding molecular substructures, enabling their use as input features for predictive models.

A radius of 2 was chosen to define the chemical environment around each atom, ensuring an appropriate balance between local and global structural features. Fingerprints were computed as 2048-bit vectors, providing a high-resolution representation of molecular structures.

These numerical encodings served as the foundation for the machine learning and deep learning workflows, acting as feature inputs for the prediction of binding affinities. The Python implementation of this step is detailed in Appendix 5.

**Generating a Deep Learning Model**

A deep learning model was developed to predict the binding affinities of small molecules targeting the ErbB receptor family. The model utilized Morgan fingerprints as features, capturing the molecular structure in a numerical format suitable for machine learning.

Morgan fingerprints served as the feature set (X), while the binding affinity values, scaled to ensure numerical stability, were used as the target labels (y). The dataset was divided into two subsets: 80% for training and 20% for testing, ensuring sufficient data for model learning and evaluation.

The deep learning model was constructed as a feedforward neural network, designed to predict binding affinity values directly. The input layer was configured to process 2048-bit Morgan fingerprint vectors, serving as molecular descriptors. Three hidden layers formed the core of the network, each employing the ReLU activation function to introduce nonlinearity and enhance learning capacity. These layers consisted of 128, 64, and 64 neurons, respectively, providing a balance between model complexity and computational efficiency. Finally, the output layer featured a single neuron dedicated to the regression task, delivering the predicted binding affinity as the model's output.

The model was trained using the Adam optimizer, a robust gradient descent algorithm. The loss function was defined as the mean squared error (MSE), capturing the squared difference between predicted and actual binding affinities. The training process was conducted over 5 epochs, with a batch size of 32, and included validation on 20% of the training data to monitor performance during training.

The performance of the model was evaluated using a set of standard metrics to assess its predictive accuracy and reliability. Mean Squared Error (MSE) was calculated to quantify the average squared difference between predicted and actual values, with a focus on penalizing larger errors. Mean Absolute Error (MAE) was also employed, offering a more intuitive measure of the average magnitude of prediction errors. Additionally, the R-squared ($R^2$) metric was determined for both the training and testing datasets, providing insight into how effectively the model captured the variability in the binding affinity data. The Python code implementing this deep learning workflow is included in Appendix 6 for reference.

Results

**Macromolecules Retrieved with Keyword "erbb"**

The search for macromolecules using the keyword "erbb" resulted in the following targets related to the ErbB receptor family shown in Table 1. Each entry includes the CHEMBL ID (target_chembl_id) and the corresponding preferred name (pref_name), describing individual receptors within the epidermal growth factor receptor (EGFR) family.

**Table 1.** Macromolecular targets related to the ErbB receptor family retrieved using the keyword "erbb."

| target_chembl_id | pref_name |
|---|---|
| CHEMBL203 | Epidermal growth factor receptor erbB1 |
| CHEMBL4523680 | Protein cereblon/Epidermal growth factor receptor |
| CHEMBL2111431 | Epidermal growth factor receptor and ErbB2 (HER1 and HER2) |
| CHEMBL2363049 | Epidermal growth factor receptor |
| CHEMBL4523998 | VHL/EGFR |
| CHEMBL5838 | Receptor tyrosine-protein kinase erbB-3 |
| CHEMBL2311234 | Receptor tyrosine-protein kinase erbB-2 |
| CHEMBL3137284 | MER intracellular domain/EGFR extracellular domain chimera |
| CHEMBL4523747 | EGFR/PPP1CA |
| CHEMBL4630723 | ErbB-2/ErbB-3 heterodimer |
| CHEMBL3848 | Receptor protein-tyrosine kinase erbB-2 |
| CHEMBL3009 | Receptor protein-tyrosine kinase erbB-4 |
| CHEMBL1824 | Receptor protein-tyrosine kinase erbB-2 |
| CHEMBL4802031 | Baculoviral IAP repeat-containing protein 2/Epidermal growth factor receptor |
| CHEMBL4106134 | FASN/HER2 |

These target_chembl_id entries formed the basis for identifying small molecule inhibitors in the subsequent step.

**Searching for Small Molecules as ErbB Inhibitors**

The second step involved querying the ChEMBL database using the target_chembl_id entries retrieved from the macromolecule search. This query aimed to identify small molecule inhibitors targeting the ErbB receptor family. The dataset was carefully filtered to include only compounds with IC50 values, ensuring the relevance of the data for evaluating inhibitor potency.

In total, 15,161 small molecule inhibitors were identified, each targeting members of the ErbB receptor family. The standard_value column contains the IC50 values, a crucial metric for assessing potency, where smaller values correspond to higher inhibitory potency. IC50 values are expressed in nanomolar (nM) units.

A representative sample of the output dataset is provided in Table 2, showcasing the inhibitors identified alongside their IC50 values and additional relevant details.

The pref_name column highlights the specific ErbB family receptors targeted by each compound, with a notable focus on Receptor protein-tyrosine kinase erbB-4. Meanwhile, the canonical_smiles column presents the molecular structure of each compound in SMILES format, facilitating downstream applications such as molecular docking and feature generation for machine learning models.

Table 2. Representative examples of small molecule inhibitors targeting the ErbB receptor family, including IC50 values and molecular structure representations.

| molecule_chembl_id | canonical_smiles | standard_value | pref_name | search_term |
|---|---|---|---|---|
| CHEMBL30432 | CN(C)c1cc2c(Nc3ccc4c(cnn4Cc4ccccc4)c3)ncnc2cn1 | 20.0 | Receptor protein-tyrosine kinase erbB-4 | erbb |
| CHEMBL545315 | C=CC(=O)Nc1cc2c(Nc3ccc(F)c(Cl)c3)ncnc2cc1OCCCN1CCOCC1.Cl.Cl | 31.0 | Receptor protein-tyrosine kinase erbB-4 | erbb |
| CHEMBL437890 | CN1CCN(CCC#CC(=O)Nc2cc3c(Nc4ccc(F)c(Br)c4)ncnc3cn2)CC1 | 0.8 | Receptor protein-tyrosine kinase erbB-4 | erbb |
| CHEMBL203599 | CN1CCN(CCC#CC(=O)Nc2cc3c(Nc4ccc(F)c(Cl)c4)ncnc3cn2)CC1 | 1.2 | Receptor protein-tyrosine kinase erbB-4 | erbb |
| CHEMBL205059 | O=C(C#CCCN1CCCCC1)Nc1cc2c(Nc3ccc(F)c(Cl)c3)ncnc2cn1 | 1.8 | Receptor protein-tyrosine kinase erbB-4 | erbb |

| CHEMBL204638 | O=C(C#CCCN1CCOCC1)Nc1cc2c(Nc3ccc(F)c(Br)c3)ncnc2cn1 | 1.0 | Receptor protein-tyrosine kinase erbB-4 | erbb |

This dataset of 15,161 small molecules with associated IC50 values will be used in the subsequent steps, including data cleaning, molecular docking simulations, and machine learning model development to predict binding affinities. The Python code for this step is included in Appendix 2.

**Cleaning Data**

The dataset obtained from Step 2, which initially contained 15,161 small molecules, was carefully processed and cleaned to ensure its suitability for further analyses. To eliminate redundancies, duplicate entries with identical molecule_chembl_id were identified and removed, retaining only the rows with the lowest standard_value, which reflects the potency of the inhibitor. This ensured that only the most relevant data, indicating higher potency, remained in the dataset.

In addition, rows with missing or invalid values in the standard_value column, such as NaN or 0, were discarded. These entries would not contribute meaningful information to the binding affinity analysis, and their removal improved the integrity of the dataset.

Irrelevant columns, specifically pref_name and search_term, were also removed, as they did not contribute to the machine learning models or molecular docking workflows that followed. Following these cleaning steps, the dataset was reduced to 9,947 unique small molecules, now properly formatted and ready for subsequent molecular docking simulations and the development of deep learning models to predict binding affinities. The Python code used for this data cleaning process is available in Appendix 3.

**Binding Affinity Calculation Using AutoDock Vina**

The binding affinities of the 9,947 small molecules, which were obtained after data cleaning in Step 3, were calculated using AutoDock Vina, a widely used molecular docking tool. The process began with ligand preparation, where each ligand was represented by its canonical SMILES string from the cleaned dataset. These SMILES strings were then converted into PDBQT files, a necessary format for AutoDock Vina docking simulations. This conversion was carried out using RDKit and Open Babel tools, with the specific Python code provided in Appendix 4.

Following the ligand preparation, molecular docking simulations were performed for each ligand using the AutoDock Vina tool. In these simulations, the receptor protein—representing the ErbB target—was docked with each ligand to predict their binding affinities. The docking results were carefully analyzed, and the best binding affinity for each ligand was extracted from the Vina output, specifically from the line marked as "REMARK VINA RESULT," which provides the optimal docking score.

The binding affinity values, measured in kcal/mol, were used to represent the strength of interaction between each ligand and its corresponding receptor. This information was then compiled into a final dataset, which was saved as a CSV file for subsequent analyses. The output dataset contains several key columns, including the molecule_chembl_id (the unique identifier for each ligand), canonical_smiles (the SMILES notation of the ligand's structure), standard_value (the IC50 value in nM, indicating the potency of the inhibitor), and binding_affinity (the predicted binding affinity in kcal/mol from the AutoDock Vina docking simulation).

A snippet of the docking output is shown in Table 3, demonstrating a few examples of the small molecules along with their corresponding IC50 values and predicted binding affinities. The receptor used for docking was an ErbB protein model, provided in PDBQT format, and the results, including the binding affinity values, were saved to the output file named erbb_docking.csv.

Table 3. Example of docking results showing small molecules, IC50 values, and predicted binding affinities.

| molecule_chembl_id | canonical_smiles | standard_value | binding_affinity |
| --- | --- | --- | --- |
| CHEMBL5073622 | C=CC(=O)N1CC[C@H](Oc2nc(-c3n[nH]c(=O)[nH]3)cc3ccccc23)C1 | 0.000000005012 | -9.893 |
| CHEMBL63786 | Brc1cccc(Nc2ncnc3cc4ccccc4cc23)c1 | 0.003 | -9.535 |
| CHEMBL35820 | CCOc1cc2ncnc(Nc3cccc(Br)c3)c2cc1OCC | 0.006 | -7.689 |
| CHEMBL66031 | Brc1cccc(Nc2ncnc3cc4[nH]cnc4cc23)c1 | 0.008 | -8.397 |
| CHEMBL53753 | CNc1cc2c(Nc3cccc(Br)c3)ncnc2cn1 | 0.008 | -7.773 |

**Morgan Fingerprints Calculation**

In this step, Morgan fingerprints were computed for each small molecule in the dataset, which includes the ligands along with their binding affinities obtained from the docking simulations in Step 4. Morgan fingerprints serve as a compact numerical representation of each molecule's structure, making them suitable for use as input features in machine learning models.

The process began with the generation of the Morgan fingerprints. For each ligand, the fingerprint was derived from its SMILES notation. The calculation utilized a radius of 2, capturing a 2-step neighborhood around each atom, and a bit length of 2048, meaning that the fingerprint is represented as a 2048-bit vector. This vector encodes various substructures within the molecule, providing a detailed representation of its structural characteristics.

Once the fingerprints were calculated, a new column titled morgan_fingerprint was added to the dataset. This column contained the calculated fingerprints for each ligand. The updated dataset now included the molecular identifiers, SMILES, IC50 values, docking binding affinities, and the newly computed Morgan fingerprints, which are essential for subsequent machine learning analyses.

The final dataset, including the molecular fingerprints, was saved in the file cleaned_erbb_docking_with_fingerprints.csv. In this dataset, the morgan_fingerprint column represents each fingerprint as a bit string, where each bit indicates the presence or absence of a specific substructure within the molecule.

This dataset is now ready to be used as input for machine learning models, such as deep learning or other regression models, to predict the binding affinities of the ligands based on their structural features.

**Model Training and Evaluation**

In this study, a deep learning model was developed to predict the binding affinity of ligands targeting the ErbB receptor family, based on their Morgan fingerprints. The model's performance was evaluated using several key metrics, including Mean Squared Error (MSE), Mean Absolute Error (MAE), and R-squared ($R^2$) values for both the training and test datasets.

To ensure that the dataset was focused on potent inhibitors, the dataset was filtered to include only ligands with binding affinity values ≤ -5 kcal/mol. The ligands' molecular structures were represented by their Morgan fingerprints, which were derived from their SMILES notations. These fingerprints were then converted into binary bit vectors. Additionally, to aid the model's convergence during training, the binding affinity values were normalized using StandardScaler.

A deep learning model was constructed using the TensorFlow Keras library. The architecture consisted of an input layer, hidden layers and output layer. Input Layer accepted the binary Morgan fingerprint as input, representing the molecular structure of the ligands. Hidden Layers contained three fully connected layers with 128, 64, and 64 neurons, respectively. The ReLU activation function was used for all hidden layers. A single neuron in the output layer was responsible for predicting the binding affinity value of the ligand, formulated as a regression task.

The model was compiled using the Adam optimizer, with the Mean Squared Error (MSE) loss function, which is commonly used for regression tasks.

The model was trained for 50 epochs, with a batch size of 32. During the training process, 20% of the data was used for validation to monitor overfitting. The model's performance was evaluated on both the training and test sets using the evaluation metrics such as Mean Squared Error (MSE) = 0.2591, Mean Absolute Error (MAE) = 0.3658, R-squared (Training Set) = 0.9389, and R-squared (Test Set) = 0.7731.

To further assess model performance, a scatter plot comparing the true binding affinity values with the predicted binding affinities was generated. The plot also includes regression lines for both the training and test sets, demonstrating the model's ability to accurately predict binding affinities.

Following training, the model was saved in the H5 format (*erbB_binding_affinity_model.h5*) for future use, including potential deployment in further studies or clinical applications. A plot comparing experimental binding affinity with predicted binding affinity for the ErbB inhibitors was generated and saved as *binding_affinity_comparison_plot.png* (Figure 2). The plot shows a strong correlation between the true and predicted values, particularly for the training set, underscoring the effectiveness of the model in predicting ligand binding affinity.

**Figure 2**. Experimental vs Predicted Binding Affinity for ErbB Inhibitors

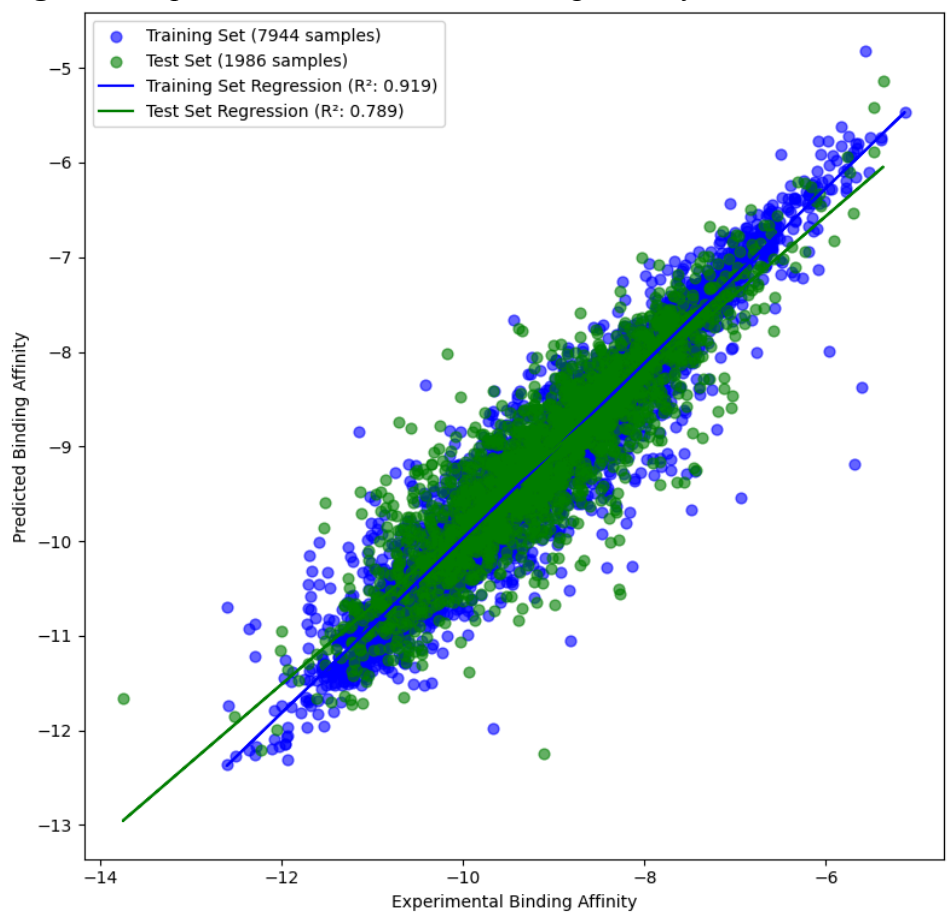

# Discussion

This study aimed to develop a deep learning model for predicting the binding affinity of ligands targeting the ErbB receptor family. Ligand data was sourced from the ChEMBL database, which provided a reliable foundation for our analysis. The dataset included compounds with known biological activity as inhibitors of ErbB. The binding affinities of each compound in interaction with ErbB was calculated using the Vina score function. This allowed us to perform downstream analyses with confidence that the data were relevant to the target prediction task.

**Strengths of the Dataset**

The ChEMBL database is widely recognized for its comprehensive and high-quality bioactivity data, making it a trustworthy source for ligand information [50]. The compounds selected for this study were well-documented, offering strong experimental validation for their interactions with the ErbB receptor. This feature of the dataset bolstered the credibility of our predictions.

**Challenges and Data Cleaning**

While the dataset was a robust starting point, it required extensive preprocessing due to missing or invalid data for some ligands (e.g., missing binding affinities). Such discrepancies in the data prompted us to clean the dataset by removing duplicates and handling invalid entries, which ensured that only reliable data were retained. After cleaning, the dataset was reduced from 15,161 to 9,947 ligands with valid IC50 values, ensuring the integrity and suitability of the data for modeling.

The data cleaning process, though necessary, reduced the sample size, which can limit the amount of training data available for model development. Despite this reduction, maintaining a dataset of high-quality data was prioritized to avoid bias and inaccuracies in the modeling process.

**Morgan Fingerprints as Molecular Representations**

In this study, we utilized Morgan fingerprints to represent the ligands' molecular structures. These fingerprints are widely used in cheminformatics because they efficiently capture the key structural features of molecules, facilitating molecular similarity searches and feature extraction. The conversion of SMILES representations into binary Morgan fingerprints allowed us to add a compact, informative feature to our dataset for training the machine learning model.

Morgan fingerprints are advantageous due to their ability to reduce dimensionality while retaining critical molecular information. However, one limitation is that they may not capture all relevant chemical features needed for accurate predictions, especially in the context of complex molecular interactions. Additionally, the choice of fingerprint parameters, such as the radius and bit size, can significantly influence the model's performance.

**Molecular Docking to Predict Binding Affinities**

To predict the binding affinities of the ligands for the ErbB receptor, we employed molecular docking simulations using AutoDock Vina. This computational tool is widely used for its efficiency and accuracy in predicting ligand-receptor interactions. The docking results provided predicted binding affinities, which were essential for training the predictive model.

Although molecular docking is a standard technique for such predictions, it has inherent challenges. The accuracy of docking results can be influenced by several factors, including the quality of the receptor structure, the selection of docking parameters, and the precision of the binding site. In this study, we assumed the receptor was adequately prepared, but in real-world applications, challenges related to docking accuracy could arise.

**Model Development and Evaluation**

The core of this study involved developing a deep learning model to predict the binding affinity of ErbB inhibitors using the Morgan fingerprints and normalized binding affinities as input features. The model was designed with three hidden layers, employing ReLU activation functions to capture complex relationships between the ligands' molecular features and their predicted binding affinity. Performance was evaluated based on R-squared ($R^2$), Mean Squared Error (MSE), and Mean Absolute Error (MAE), metrics that provide insights into the model's accuracy [51].

**Performance Analysis**

The deep learning model showed strong performance on the training set, with an $R^2$ value of 0.939, indicating that the model captured most of the variance in the training data. On the test set, the $R^2$ value dropped to 0.773, which suggests that the model generalizes reasonably well to unseen data. The performance on the test set is still strong, demonstrating that the model can make reliable predictions on new data. However, the drop in performance also indicates room for improvement, particularly in enhancing generalization.

The MSE of 0.259 suggests that, on average, the squared differences between predicted and actual binding affinities are relatively small. However, MSE is sensitive to outliers, and a few extreme errors could have contributed to this value. To further improve the model, addressing these outliers or refining the model's handling of complex molecular interactions could be beneficial.

With an MAE of 0.366, the model's average prediction error is relatively small, indicating that the model's predictions are close to the actual binding affinity values. MAE is less sensitive to outliers than MSE, making it a useful metric for gauging overall prediction accuracy. While the MAE suggests that the model is robust, further refinements could reduce the errors.

The high $R^2$ value of 0.939 for the training set indicates that the model explained a significant portion of the variance in the binding affinities. This suggests that the features, particularly the Morgan fingerprints, were highly informative in learning the relationships between molecular structures and binding affinity.

The R² value of 0.773 for the test set suggests that the model generalizes well to new data, although the decrease from the training set indicates that further improvement is possible. The slight drop in R² between the training and test sets is a common occurrence in machine learning, especially when the model is complex or trained on a limited dataset.

**Conclusion**

In conclusion, the deep learning model developed in this study offers a promising approach for predicting the binding affinity of ErbB inhibitors based on their molecular fingerprints. The results indicate that the model can accurately predict binding affinities, as reflected in the evaluation metrics. The Mean Squared Error (0.259) and Mean Absolute Error (0.366) suggest that the model's predictions are relatively close to the actual binding affinities, while the high R-squared values for both the training (0.939) and test sets (0.773) demonstrate the model's ability to explain a significant portion of the variance in binding affinity.

Although the model shows strong performance, the drop in R-squared from the training set to the test set highlights an opportunity for improvement, particularly in enhancing the model's generalization capabilities. Potential improvements can be achieved by better handling outliers, incorporating regularization techniques, and expanding the feature set. These refinements will further increase the model's robustness and its applicability to drug discovery and the prediction of molecular interactions.

This study underscores the potential of deep learning models in predicting molecular properties and provides a solid foundation for future work aimed at improving predictive accuracy. With further refinements, the model could be a valuable tool in accelerating the drug discovery process, particularly in the context of ErbB-targeted therapies.

**Acknowledgments**

This research was supported by the LPPM UNG through the PNBP-Universitas Negeri Gorontalo, under DIPA-UNG No. 653/UN47/HK.02/2024, as part of the Riset Unggulan Fakultas program, with contract No. B/812/UN47.D1.1/PT.01.03/2024. We would like to express our heartfelt gratitude to the Fugaku Supercomputer facility for providing the computational resources essential for performing the molecular docking simulations in this study. The high-performance computing power of Fugaku enabled the efficient and precise calculation of binding affinities for the ligands targeting the ErbB receptor family. Without their invaluable support, this research would not have been feasible. Furthermore, we extend our thanks to the ChEMBL database for offering the reliable data that served as the foundation for this study.

**Conflicts of interest statement**

All the authors declare no conflicts of interest.

**Authors' contributions**

All the authors had equally contributed to the current study. All authors read and approved the finalized version.

Appendix 1. Searching Macromolecules with keywords

```python
import os
import numpy as np
import pandas as pd
from chembl_webresource_client.new_client import new_client
import tkinter as tk

def search_targets(query_list, job_dir, result_text):
    target_list = []
    for search_term in query_list:
        np.random.seed(1)

        # Load target data
        target = new_client.target

        # Perform target search
        target_query = target.search(search_term)
        targets = pd.DataFrame.from_dict(target_query)

        # Select columns and include search_term
        search_term_formatted = search_term.replace(" ", "-")
        selected_columns = ['target_chembl_id', 'pref_name']
        targets['search_term'] = search_term_formatted
        targets = targets[selected_columns]

        # Save DataFrame to CSV
        targets_output_file_path = os.path.join(job_dir, f'targets_{search_term_formatted}.csv')
        targets.to_csv(targets_output_file_path, index=False)

        # Cetak ke result_text
        if result_text is not None:
            result_text.insert(tk.END, f"{search_term.upper()}'s targets saved into {targets_output_file_path}\n")
```

```python
            target_list.append(targets_output_file_path)

    return target_list

def main():
    # Definisikan query_list langsung di dalam kode
    query_list = ['erbb']  # Gantilah ini dengan kata kunci yang diinginkan

    job_dir = os.getcwd()

    # Membuat GUI menggunakan tkinter
    root = tk.Tk()
    root.title("ChEMBL Target Search")

    # Membuat komponen teks untuk menampilkan hasil
    result_text = tk.Text(root, height=10, width=50)
    result_text.pack()

    # Menjalankan pencarian target
    search_targets(query_list, job_dir, result_text)

    # Menampilkan GUI
    root.mainloop()

if __name__ == "__main__":
    main()
```

Appendix 2. Searching Small Molecules as ErbB Inhibitors

```python
import pandas as pd
import numpy as np
import os
from chembl_webresource_client.new_client import new_client
from rdkit import Chem
from tqdm import tqdm
from concurrent.futures import ProcessPoolExecutor
from glob import glob
from tkinter import ttk
import tkinter as tk

# Function to validate smiles
def validate_smiles(smiles):
    mol = Chem.MolFromSmiles(smiles)
    if mol is not None:
        return Chem.MolToSmiles(mol)
    else:
        return None

def combine_csv_files(target, target_dir):
    # Read all CSV files in the directory
    csv_files = glob(os.path.join(target_dir, '*.csv'))

    if not csv_files:
        print("No CSV files found in the directory.")
        return None  # Return None if no files found

    # Read each file and combine them into one DataFrame
    dfs = [pd.read_csv(file) for file in csv_files if os.path.getsize(file) > 0]

    if not dfs:
        print(f"No data in CSV files in the directory {target_dir}.")
        return None  # Return None if no data in files

    combined_df = pd.concat(dfs, ignore_index=True)
    output_file = os.path.join(target_dir, f'{target}.csv')

    # Save the combined DataFrame to one CSV file
    combined_df.to_csv(output_file, index=False)
    print(f"Data {target_dir} saved in {output_file}")
```

```python
    return combined_df

def process_chembl_id(chembl_id, target, target_dir, other, others_column):
    try:
        # Load data
        activity = new_client.activity
        res = activity.filter(target_chembl_id=chembl_id).filter(standard_type="IC50")
        df = pd.DataFrame.from_dict(res)

        # Extract columns
        if 'molecule_chembl_id' in df.columns:
            mol_cid = list(df.molecule_chembl_id)
        else:
            mol_cid = []

        if 'canonical_smiles' in df.columns:
            canonical_smiles = list(df.canonical_smiles)
        else:
            canonical_smiles = []

        if 'standard_value' in df.columns:
            standard_value = list(df.standard_value)
        else:
            standard_value = []

        # Create DataFrame with all columns
        data_tuples = list(zip(mol_cid, canonical_smiles, standard_value))
        df = pd.DataFrame(data_tuples, columns=['molecule_chembl_id', 'canonical_smiles', 'standard_value'])

        # Clean and remove duplicates
        df = df.dropna(subset=['molecule_chembl_id'])
        df = df.dropna(subset=['canonical_smiles'])
        df['canonical_smiles'] = df['canonical_smiles'].apply(validate_smiles)
        df = df.drop_duplicates(subset=['molecule_chembl_id'], keep='first', ignore_index=True)

        # Add a new column others and 'target' with the value {target}
        df[others_column] = other
        df['search_term'] = target

        output_file_path = os.path.join(target_dir, f'{chembl_id}.csv')
        df.to_csv(output_file_path, index=False)
```

```python
        return f"Processed {chembl_id}"

    except Exception as e:
        return f"Error processing {chembl_id}: {str(e)}"

def search_molecules(job_dir):
    # Set the random seed for reproducibility
    np.random.seed(1)

    df_all_target = []
    targets = []
    all_results = []

    targets_files = glob(os.path.join(job_dir, 'targets_*.csv'))

    for targets_file_path in targets_files:
        df_targets = pd.read_csv(targets_file_path)
        target_chembl_ids = df_targets['target_chembl_id']
        others_columns = df_targets.columns.difference(['target_chembl_id'])  # Get columns other than 'target_chembl_id'
        target = os.path.splitext(os.path.basename(targets_file_path))[0].replace("targets_", "")
        target_dir = os.path.join(job_dir, target)
        targets.append(target)

        if not os.path.exists(target_dir):
            os.mkdir(target_dir)

        max_workers = os.cpu_count()
        futures = []

        with ProcessPoolExecutor(max_workers=max_workers) as executor:
            # Process chembl_ids in parallel
            for chembl_id in target_chembl_ids:
                other_values = df_targets[df_targets['target_chembl_id'] == chembl_id][others_columns].iloc[0]

                future = executor.submit(process_chembl_id, chembl_id, target, target_dir, other_values, others_columns)
                futures.append(future)

            # Wait for all tasks to complete
            results = [future.result() for future in futures]
```

```python
        combined_df = combine_csv_files(target, target_dir)
        print(combined_df)

        if not combined_df.empty:  # Check if DataFrame is not empty before appending
            df_all_target.append(combined_df)
            # Remove the following lines as DataFrame doesn't have 'head' and 'info' methods
            print(df_all_target)

    if df_all_target:
        # Combine all DataFrames in the list
        df_all_target_combined = pd.concat(df_all_target, ignore_index=True)

        # Save the combined DataFrame to one CSV file
        csv_molecules = os.path.join(job_dir, 'merging_' + '_'.join(targets) + '.csv')
        df_all_target_combined.to_csv(csv_molecules, index=False)
        if result_text is not None:
            result_text.insert("end", f"Data all targets saved in {csv_molecules}\n")
        return csv_molecules

    return csv_molecules

if __name__ == "__main__":
    job_dir = os.getcwd()
    search_molecules(job_dir)
```

Appendix 3. Cleaning Data

```python
import pandas as pd

def clean_combined_data(input_file, output_file):
    """
    Cleans the combined CSV data by:
    1. Removing duplicate rows based on 'molecule_chembl_id',
       keeping the row with the lowest 'standard_value'.
    2. Dropping rows where 'standard_value' is NaN or 0.
    3. Dropping the columns 'pref_name' and 'search_term'.

    Args:
        input_file (str): Path to the input CSV file.
        output_file (str): Path to save the cleaned CSV file.

    Returns:
        str: Path to the cleaned file.
    """
    # Load the CSV file into a DataFrame
    df = pd.read_csv(input_file)

    # Drop rows where 'standard_value' is NaN or 0
    if 'standard_value' in df.columns:
        df['standard_value'] = pd.to_numeric(df['standard_value'], errors='coerce')  # Ensure standard_value is numeric
        df = df[df['standard_value'].notna() & (df['standard_value'] != 0)]  # Remove rows where standard_value is NaN or 0

    # Drop duplicates by molecule_chembl_id, keeping the row with the lowest standard_value
    df = df.sort_values('standard_value').drop_duplicates(subset='molecule_chembl_id', keep='first')

    # Drop the columns 'pref_name' and 'search_term' if they exist
    columns_to_drop = ['pref_name', 'search_term']
    df = df.drop(columns=[col for col in columns_to_drop if col in df.columns])

    # Save the cleaned DataFrame to a new CSV file
    df.to_csv(output_file, index=False)
    print(f"Cleaned data saved to: {output_file}")
    return output_file

# Example usage
```

```python
if __name__ == "__main__":
    input_file = "erbb_docking.csv"  # Path to the input file
    output_file = "cleaned_erbb_docking.csv"  # Path to save the cleaned file

    clean_combined_data(input_file, output_file)
```

Appendix 4. Binding Affinity Calculation Using Autodock Vina

```python
import os
import pandas as pd
from rdkit import Chem
from rdkit.Chem import AllChem
from subprocess import run, DEVNULL, CalledProcessError
from tqdm import tqdm
from concurrent.futures import ProcessPoolExecutor

def smiles_to_pdbqt(smiles, ligand_name):
    try:
        mol = Chem.MolFromSmiles(smiles)
        if mol is None:
            raise ValueError(f"Invalid SMILES: {smiles}")

        mol = Chem.AddHs(mol)
        AllChem.EmbedMolecule(mol, AllChem.ETKDG())
        AllChem.UFFOptimizeMolecule(mol)

        pdb_path = f"{ligand_name}.pdb"
        Chem.MolToPDBFile(mol, pdb_path)

        pdbqt_path = f"{ligand_name}.pdbqt"
        run(["obabel", pdb_path, "-O", pdbqt_path, "--gen3d"], stdout=DEVNULL, stderr=DEVNULL, check=True)

        os.remove(pdb_path)  # Hapus file PDB setelah konversi
        return pdbqt_path
    except (ValueError, CalledProcessError, Exception) as e:
        print(f"Error converting {ligand_name} to PDBQT: {e}")
        return None

def run_vina_docking(receptor, config_file, ligand_pdbqt, ligand_name):
    try:
        output_pdbqt = f"{ligand_name}_out.pdbqt"

        run([
            "vina",
            "--receptor", receptor,
            "--ligand", ligand_pdbqt,
            "--config", config_file,
```

```python
            "--out", output_pdbqt
        ], stdout=DEVNULL, stderr=DEVNULL, check=True)

        return output_pdbqt
    except CalledProcessError as e:
        print(f"Vina docking failed for {ligand_name}: {e}")
        return None

def extract_best_binding_affinity(output_pdbqt):
    try:
        with open(output_pdbqt, 'r') as file:
            for line in file:
                if "REMARK VINA RESULT" in line:
                    return float(line.split()[3])  # Mengambil nilai affinity terbaik
    except Exception as e:
        print(f"Error reading output file {output_pdbqt}: {e}")
    return None

def process_ligand(row, receptor, config_file):
    smiles = row['canonical_smiles']
    molecule_id = row['molecule_chembl_id']
    standard_value = row['standard_value']
    ligand_name = f"ligand_{molecule_id}"

    ligand_pdbqt = smiles_to_pdbqt(smiles, ligand_name)
    if not ligand_pdbqt:
        return molecule_id, smiles, standard_value, None

    output_pdbqt = run_vina_docking(receptor, config_file, ligand_pdbqt, ligand_name)
    best_affinity = None
    if output_pdbqt:
        best_affinity = extract_best_binding_affinity(output_pdbqt)
        os.remove(output_pdbqt)

    if os.path.exists(ligand_pdbqt):
        os.remove(ligand_pdbqt)

    return molecule_id, smiles, standard_value, best_affinity

def perform_docking(input_file, receptor, config_file, output_csv):
    if not os.path.exists(receptor):
```

```python
        print(f"Receptor file not found: {receptor}")
        return

    if not os.path.exists(config_file):
        print(f"Config file not found: {config_file}")
        return

    input_df = pd.read_csv(input_file)
    if 'canonical_smiles' not in input_df.columns or 'molecule_chembl_id' not in input_df.columns or 'standard_value' not in input_df.columns:
        print("Required columns 'canonical_smiles', 'molecule_chembl_id', or 'standard_value' missing in input file.")
        return

    # Pastikan file output CSV memiliki header jika belum ada
    if not os.path.exists(output_csv):
        with open(output_csv, 'w') as f:
            f.write("molecule_chembl_id,canonical_smiles,standard_value,binding_affinity\n")

    # Proses secara paralel
    rows = input_df.to_dict('records')
    with ProcessPoolExecutor(max_workers=1) as executor:
        futures = [executor.submit(process_ligand, row, receptor, config_file) for row in rows]

        for future in tqdm(futures, total=len(rows), desc="Docking ligands"):
            try:
                result = future.result()
                if result:
                    molecule_id, smiles, standard_value, best_affinity = result
                    # Tambahkan hasil ke file CSV output
                    with open(output_csv, 'a') as f:
                        f.write(f"{molecule_id},{smiles},{standard_value},{best_affinity}\n")
            except Exception as e:
                print(f"Error during docking: {e}")

    print(f"Docking results saved to {output_csv}")

if __name__ == "__main__":
    input_file = "cleaned_erbb.csv"
    receptor = "rec_erbb.pdbqt"
    config_file = "config.txt"
    output_csv = "erbb_docking.csv"
```

perform_docking(input_file, receptor, config_file, output_csv)

Appendix 5. Morgan Fingerprint Calculation of Ligand

```python
import pandas as pd
from rdkit import Chem
from rdkit.Chem import AllChem

# Load the dataset
df = pd.read_csv('cleaned_erbb_docking.csv')

# Function to generate Morgan Fingerprint from SMILES
def generate_morgan_fingerprint(smiles, radius=2, n_bits=2048):
    mol = Chem.MolFromSmiles(smiles)
    if mol:
        # Generate Morgan fingerprint with given radius and number of bits
        fingerprint = AllChem.GetMorganFingerprintAsBitVect(mol, radius, nBits=n_bits)
        return fingerprint.ToBitString()  # Return fingerprint as bit string
    else:
        return None

# Apply the function to each molecule in the dataset
df['morgan_fingerprint'] = df['canonical_smiles'].apply(generate_morgan_fingerprint)

# Save the dataset with the new 'morgan_fingerprint' column
df.to_csv('cleaned_erbb_docking_with_fingerprints.csv', index=False)

# Display first few rows of the dataframe
print(df.head())
```

Appendix 5. Generate deep learning model

```python
import pandas as pd
import numpy as np
from sklearn.model_selection import train_test_split
from sklearn.preprocessing import StandardScaler
from sklearn.metrics import mean_squared_error, r2_score
import tensorflow as tf
from tensorflow.keras import layers, models
import matplotlib.pyplot as plt
from sklearn.linear_model import LinearRegression

# Load the dataset
df = pd.read_csv('cleaned_erbb_docking_with_fingerprints.csv')

# Filter rows where binding_affinity <= -5
df_filtered = df[df['binding_affinity'] <= -5]

# Prepare features (Morgan fingerprints) and labels (binding_affinity)
# Convert Morgan fingerprint from bitstring to a numpy array of integers
def fingerprint_to_array(fingerprint):
    return np.array([int(bit) for bit in fingerprint])

# Extract features and labels
X = np.array(df_filtered['morgan_fingerprint'].apply(fingerprint_to_array).tolist())
y = df_filtered['binding_affinity'].values

# Normalize the binding affinity values
scaler = StandardScaler()
y_scaled = scaler.fit_transform(y.reshape(-1, 1)).flatten()  # Normalize the target (binding affinity)

# Split the dataset into training and testing sets
X_train, X_test, y_train, y_test = train_test_split(X, y_scaled, test_size=0.2, random_state=42)

# Print the number of samples in training and test sets
num_train_samples = len(X_train)
num_test_samples = len(X_test)
print(f'Number of molecules in Training Set: {num_train_samples}')
print(f'Number of molecules in Test Set: {num_test_samples}')

# Build the deep learning model
model = models.Sequential()
```

```python
model.add(layers.InputLayer(input_shape=(X_train.shape[1],)))  # Input layer
model.add(layers.Dense(128, activation='relu'))  # First hidden layer
model.add(layers.Dense(64, activation='relu'))   # Second hidden layer
model.add(layers.Dense(64, activation='relu'))   # Third hidden layer
model.add(layers.Dense(1))  # Output layer (regression)

# Compile the model
model.compile(optimizer='adam', loss='mean_squared_error', metrics=['mae'])

# Train the model
history = model.fit(X_train, y_train, epochs=50, batch_size=32, validation_split=0.2, verbose=1)

# Predict on training and test data
y_pred_train_scaled = model.predict(X_train)
y_pred_test_scaled = model.predict(X_test)

# Inverse transform the predicted values to the original scale
y_pred_train = scaler.inverse_transform(y_pred_train_scaled.reshape(-1, 1)).flatten()
y_pred_test = scaler.inverse_transform(y_pred_test_scaled.reshape(-1, 1)).flatten()

# Inverse transform the true values to the original scale
y_train_original = scaler.inverse_transform(y_train.reshape(-1, 1)).flatten()
y_test_original = scaler.inverse_transform(y_test.reshape(-1, 1)).flatten()

# Calculate R-squared for both train and test sets
r2_train = r2_score(y_train_original, y_pred_train)
r2_test = r2_score(y_test_original, y_pred_test)

# Calculate evaluation metrics
mse = mean_squared_error(y_test_original, y_pred_test)
mae = np.mean(np.abs(y_test_original - y_pred_test))

print(f"Mean Squared Error: {mse}")
print(f"Mean Absolute Error: {mae}")
print(f"R-squared (Training Set): {r2_train}")
print(f"R-squared (Test Set): {r2_test}")

# Optionally, save the trained model
model.save('erbB_binding_affinity_model.h5')

# Create LinearRegression models for the training and test sets to plot regression lines
train_lr = LinearRegression()
test_lr = LinearRegression()
```

```python
# Train linear models
train_lr.fit(y_train_original.reshape(-1, 1), y_pred_train)
test_lr.fit(y_test_original.reshape(-1, 1), y_pred_test)

# Predict using the linear models
train_line = train_lr.predict(y_train_original.reshape(-1, 1))
test_line = test_lr.predict(y_test_original.reshape(-1, 1))

# Plotting the comparison between true and predicted values (combined for train and test)
plt.figure(figsize=(8, 8))

# Scatter plot for training set and test set
plt.scatter(y_train_original, y_pred_train, color='blue', alpha=0.6, label=f'Training Set ({num_train_samples} samples)', s=40)
plt.scatter(y_test_original, y_pred_test, color='green', alpha=0.6, label=f'Test Set ({num_test_samples} samples)', s=40)

# Add the regression line for training set
plt.plot(y_train_original, train_line, color='blue', linestyle='-', label=f'Training Set Regression (R²: {r2_train:.3f})')

# Add the regression line for test set
plt.plot(y_test_original, test_line, color='green', linestyle='-', label=f'Test Set Regression (R²: {r2_test:.3f})')

# Add labels and title with R-squared values in the title
plt.title(f'Experimental vs Predicted Binding Affinity for ErbB Inhibitors')
plt.xlabel('Experimental Binding Affinity')
plt.ylabel('Predicted Binding Affinity')

# Show legend
plt.legend()

# Save the plot to a file (PNG format)
plt.tight_layout()
plt.savefig('binding_affinity_comparison_plot.png')  # Save as PNG

# Optionally, save the plot in other formats like PDF, SVG, etc.
# plt.savefig('binding_affinity_comparison_plot.pdf')  # Save as PDF

# Show the plot
plt.show()
```